\documentclass[submission,copyright,creativecommons]{eptcs}
\providecommand{\event}{TTC 2013}
\usepackage[english]{babel}
\usepackage[latin1]{inputenc}	
\usepackage[T1]{fontenc}
\usepackage{graphicx}
\usepackage{url}
\usepackage{llncsdoc}
\usepackage{tabularx}
\usepackage{amsmath}
\usepackage{amssymb}
\usepackage{enumerate}
\usepackage{scrtime}
\usepackage{listings}
\usepackage{subfigure}
\usepackage{hyperref}

\hypersetup{
 pdfauthor={Georg Hinkel},
 pdftitle={NMF Solution for Flowgraph case},
 pdfsubject={TTC2013},
 pdfkeywords={MDSD,MDE}
}

\title{An \textsc{NMF} solution for the \textit{Petri Nets to State Charts} case study at the TTC 2013}
\author{Georg Hinkel
\institute{Karlsruhe Institute of Technology \\ Karlsruhe, Germany}
\email{georg.hinkel@student.kit.edu}
\and
Thomas Goldschmidt
\institute{ABB Corporate Research \\ Ladenburg, Germany}
\email{thomas.goldschmidt@de.abb.com}
\and
Lucia Happe
\institute{Karlsruhe Institute of Technology \\ Karlsruhe, Germany}
\email{lucia.kapova@kit.edu}
}
\def\titlerunning{An \textsc{NMF} solution for the \textit{Petri Nets to State Charts} case study at the TTC 2013}
\def\authorrunning{G. Hinkel, T. Goldschmidt \& L. Happe}

\begin{document}
\selectlanguage{english}
\maketitle


\begin{abstract}
Software systems are getting more and more complex. Model-driven engineering (MDE) offers ways to handle such increased complexity by lifting development to a higher level of abstraction. A key part in MDE are transformations that transform any given model into another. These transformations are used to generate all kinds of software artifacts from models. However, there is little consensus about the transformation tools. Thus, the Transformation Tool Contest (TTC) 2013 aims to compare different transformation engines. This is achieved through three different cases that have to be tackled. One of these cases is the Petri Net to State Chart case. A solution has to transform a Petri Net to a State Chart and has to derive a hierarchical structure within the State Chart. This paper presents the solution for this case using NMF Transformations as transformation engine.
\end{abstract}


\section{Introduction}
\label{ch:Introduction}

The challenge of the Petri Nets to State Charts case of the TTC 2013 \cite{ttcPN2SC} is to transform a Petri Net into a State Chart. Furthermore, both the Petri Net and the State Chart are afterwards to be reduced. This paper presents a solution for these tasks using NMF\footnote{\url{http://nmf.codeplex.com}}, an open source project to support model-driven engineering on the .NET platform for the transformation part. The reduction of Petri Net and State Charts has been written in general purpose C\# that is embedded in the transformation and makes use of the trace model created by NMF Transformations. The solution is available on SHARE\footnote{\url{http://is.ieis.tue.nl/staff/pvgorp/share/?page=ConfigureNewSession&vdi=XP-TUe_TTC13::NMF_TTC13::NMF_updated_NMF_LiveContest.vdi}}.

\section{.NET Modelling Framework (\textsc{NMF})}
\label{ch:NMF}

The .NET Modelling Framework is an open source project that provides support for model-driven software development on the .NET platform. An essential part is the model transformation engine, \textsc{NMF Transformations}, which allows to write rule-based transformations in arbitrary .NET languages using an internal DSL \cite{fowler2010domain}. The reason to implement the transformation language as internal DSL is mainly that transformation languages ought to be Turing complete \cite{sendall2003model} and thus, many advantages of external DSLs attenuate. An internal DSL, however, can make use of features of its host language. Developers used to this language feel familiar with the DSL.

\textsc{NMF Transformations} makes it possible to specify model transformations directly in C\#. For this purpose, \textsc{NMF Transformations} has a simple abstract syntax but hides the complexity in the attributes of the metaclasses which are representing functions. These functions can be specified with general purpose code that contain code as sophisticated as required. Although the transformation language might seem quite verbose, especially when compared with external model transformation languages, C\# has been chosen as host language to make it easier to write and thus maintain these transformations for C\# developers.

Currently, \textsc{NMF} does not contain a metamodeling foundation, e.g. based on MOF. Instead, \textsc{NMF Transformations} uses the concepts of the CLR (the virtual machine used on the .NET platform) to represent models and operates on plain objects (POCOs). Thus, we used an interop component to EMF, which generates classes from an Ecore metamodel. Furthermore, there exists a serializer component to load and store simple models (models that do not have references to other files).

Beside \textsc{NMF}, to the best of our knowledge, there is hardly any framework that supports model-driven engineering on the .NET platform. Microsoft offers a Visualization and Modeling SDK\footnote{\url{http://archive.msdn.microsoft.com/vsvmsdk}} for Visual Studio, which is a tool for graphical editors, and T4 as Text-To-Text-Transformation engine. However, although T4 is included in Visual for many years now, there is still hardly support for editing T4 templates. There is an add-in providing syntax-highlighting and code-completion, but still the support is much less than for writing normal e.g. C\# code. Furthermore, T4 has some restrictions like no inheritance is allowed within a T4-template. These restrictions and the lack of out-of-the-box tool support make model-driven development hard. \textsc{NMF Transformations} makes it possible to use the full tool support for C\# also for model transformations.

\section{Solution}
\label{ch:Solution}

As the case description was divided into several parts of the transformation, this section is also divided in several subsections each presenting the solution of a subtask within the transformation. Thus, section \ref{ch:Solution:Initialization} presents the initialization before section \ref{ch:Solution:Reduction} explains the reduction.

\subsection{Initialization}
\label{ch:Solution:Initialization}

The task of the initialization is to create an initial structure for the statechart model. Although it is not mentioned in the description, the first rule for this initialization is that for a Petri Net, a Statechart has to be created with a topstate, which is an AND element.

A M2M-transformation in \textsc{NMF Transformations} is specified through transformation rules, which are represented by classes. These transformation rules may only be called once per input arguments within a transformation context, i.e. a transformation pass. Furthermore, transformation rules may define dependencies to other rules. These dependencies are necessary to set, in order to let \textsc{NMF} Transformations know which rules to call and to derive the inputs of these rules. \textsc{NMF Transformations} operates on plain CLR objects and therefore does not know the structure of the metamodel. Thus the structure of the domain model has to be reflected in the dependencies of the transformation rules.

\begin{figure}[htpb]
	\centering
		\includegraphics[width=0.80\textwidth]{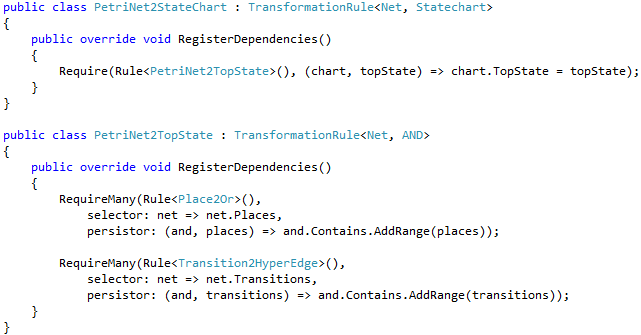}
	\caption{The transformation rules to transform the Petri Net}
	\label{fig:pnetTransformation}
\end{figure}

Thus, we have to specify that whenever a Petri Net is transformed, also all places of that Petri Net have to be transformed, before or after the transformation of the Petri Net. The only rule that is actually called by \textsc{NMF Transformations} automatically is the rule that matches the transformation request to transform a Petri Net into a state chart. This rule is depicted in Figure \ref{fig:pnetTransformation}.

The first rule in Figure \ref{fig:pnetTransformation}, \emph{PetriNet2StateChart}, is a rule that transforms a Petri Net into a State Chart. The only thing that happens is that \emph{PetriNet2TopState} is called and the resulting top state is applied to the State Chart. The second rule, \emph{PetriNet2TopState} creates this top state for each Petri Net. This rule already contains information for the next two rules: For every Place within the PetriNet, a corresponding OR element should be created and added to the initial AND top state. Furthermore, any transition should be transformed to a HyperEdge. Note that within this rule, there is no information about how the places are to be transformed, it is only required that they are transformed using the \emph{Place2OR}-rule. The same applies for the transitions. The \emph{Transition2HyperEdge} rule again requires the incoming and outgoing places to be transformed. However, the engine takes care that any place is only transformed once.

There can be multiple parameters specified to filter the dependencies, select the inputs for the dependent transformation rule or persist the output. The code example in Figure \ref{fig:pnetTransformation} already contains several cases. In the require dependency of the \textit{PetriNet2StateChart} rule, there is only a persistor specified. This persists the output of the dependent rule back to the output of the current rule, i.e. the top state of the State Chart is set. The first \emph{RequireMany} statement further has a selector specified. This selector chooses the input for the dependent transformation rule, i.e. it specifies the places that should serve as inputs for the \textit{Place2Or} transformation rule. The last dependency works in the same way.

The initialization rules can directly be reflected in transformation rules. However, due to space limitations, they are not displayed here. In total, the initialization consists of just five transformation rules:
\begin{itemize}
	\item {\textit{PetriNet2StateChart} that transforms a Petri Net into a State Chart}
	\item {\textit{PetriNet2TopState} that creates the top state for a State Chart}
	\item {\textit{Place2Basic} that creates the Basic element for each place in the Petri Net}
	\item {\textit{Place2Or} that creates the OR element for each place in the Petri Net}
	\item {\textit{Transition2HyperEdge} that creates the HyperEdge for a each transition}
\end{itemize}

With these five transformation rules, the initialization task is completed. The demanded equivalence function is implicitly stored in the transformation context, as it contains a trace where we just need to resolve a place using the \emph{Place2Basic} or \emph{Place2OR} rule. However, this trace functionality is not serialized by default. If we wanted to serialize the trace into a file, we would have to do this on our own. 

\subsection{Reduction}
\label{ch:Solution:Reduction}

As \textsc{NMF Transformations} only checks application conditions for transformation rules once, we implemented the reduction in general purpose code in C\#. However, this general purpose code is embedded in the transformation and makes use of the transformation engine, especially of the provided trace functionality.

Therefore, we extend the transformation with the given reduction rules and include the reduction code. As the rules apply on transitions, we write this code in the \textit{Transform}-method of the \emph{Transition2HyperEdge}-rule. We can just call the reduction rule from within the \textit{Transform}-method, as both the reduction rules and the transformation is written in C\#. We can even put the helper methods that we need into the class representing the \emph{Transition2HyperEdge}-rule. 

The implementations of the two main reduction rules are both placed within a single method. Both methods consist of the following steps: $(1)$ At first, it is checked whether the reduction rule is applicable at all. If not, the method immediately returns. $(2)$ Next, the rule is applied to the Petri Net and $(3)$ to the State Chart model. $(4)$ Finally, the rule checks whether another rule is applicable with new inputs. Besides step $(3)$, all these steps are implemented using traditional straight-forward imperative code that operates on the in-memory model representations. The only point where the reduction needs an access to the correspondence established by the model transformation is when the reduction rules are applied to the State Chart model. To have an access to the trace functionality, both methods have an additional parameter that is the transformation context. 

\begin{figure}[htb]
	\centering
		\includegraphics{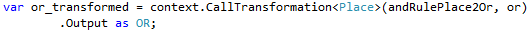}
	\caption{Calling a transformation rule within the reduction}
	\label{fig:AND_callTransformation}
\end{figure}

As the reduction rules rely on the trace functionality, it is necessary to use update the trace when new OR compounds have to be created. This is done via the code snippet shown in Figure \ref{fig:AND_callTransformation} where \textit{andRulePlace2Or} is a cached reference to an additional transformation rule that transforms a place into an OR compound without creating a Basic element.

\begin{figure}[htb]
	\centering
		\includegraphics{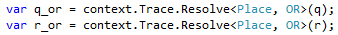}
	\caption{Using the trace functionality to resolve single places in the OR rule}
	\label{fig:OR_resolve}
\end{figure}

To query the correspondence established by the tracing functionality, we can simply use the trace component that is given as a reference of the transformation context and either resolve many places at once (as needed for the AND rule) or resolve single places (as needed for the OR rule). This tracing functionality is shown in the Figure \ref{fig:OR_resolve}.

Figure \ref{fig:OR_resolve} shows the tracing functionality without the specification of a transformation rule acting as key. Thus, the trace looks up all transformations from a Place to an OR, i.e. collects transformations using either \emph{AndRulePlace2Or} or \emph{Place2Basic}.

\section{Analysis}
\label{ch:validation}

\subsection{Performance measurements}

The average execution times on SHARE\footnote{\url{http://is.ieis.tue.nl/staff/pvgorp/share/?page=ConfigureNewSession&vdi=XP-TUe_TTC13::NMF_TTC13::NMF_updated_NMF_LiveContest.vdi}} for the performance test cases are shown in table \ref{tab:ExecutionTimesOfThePerformanceTestCases}. Each test case was repeated five times. All the tests were created in a single batch mode to delete the influence that the code has to be compiled by the JIT compiler when the transformation is run for the first time. However, the transformation has not been reused, although \textsc{NMF Transformations} allows such procedure. Instead, the transformation has been initialized every time from scratch.

The transformation used the original provided performance test cases with id-referencing scheme and thus, the load times are a bit slow.

\subsection{Profiler results}

As \textsc{NMF Transformations} is used through an internal DSL for C\#, the tool support integrated within Visual Studio and other tools can be applied to model transformations created with it. Besides a great debugging and refactoring support, this means that a profiler can be applied to the model transformations. This made it possible to unveil the biggest performance degrade in previous versions. These previous versions used ordered sets due to the index-based referencing scheme in the test models. As the profiler showed, the remove operation of these ordered sets was consuming large shares of the execution time. Indeed, by changing the underlying data structure for the collections, the performance increased by two magnitudes.

\section{Conclusion}
\label{ch:conclusion}

In this paper we have presented a solution to the TTC 2013 \textit{Petri Nets to State Charts} case based on \textsc{NMF Transformations}. It was not possible to support every bit of the transformation with \textsc{NMF}, as the reduction was entirely written in general purpose code interacting with the transformation engine. However, it was easy to integrate the general purpose code into the transformation. 

We suggest the high points of our solution as 
\begin{itemize}
	\item {\textbf{Good execution speed}. The biggest test model is transformed in a few seconds.}
	\item {\textbf{Easy integration of general purpose code}. The whole reduction part is written as general purpose code that interacts with the transformation.}
	\item {\textbf{Great tool support}, as \textsc{NMF Transformations} can reuse for example debugging, profiling, refactoring, testing and continuous integration support for C\#.}
\end{itemize}

\phantomsection
\addcontentsline{toc}{chapter}{\bibname}

\bibliographystyle{eptcs}
												  

\bibliography{ttc2013_pn2sc}

\begin{thebibliography}{1}
\providecommand{\bibitemdeclare}[2]{}
\providecommand{\surnamestart}{}
\providecommand{\surnameend}{}
\providecommand{\urlprefix}{Available at }
\providecommand{\url}[1]{\texttt{#1}}
\providecommand{\href}[2]{\texttt{#2}}
\providecommand{\urlalt}[2]{\href{#1}{#2}}
\providecommand{\doi}[1]{doi:\urlalt{http://dx.doi.org/#1}{#1}}
\providecommand{\bibinfo}[2]{#2}

\bibitemdeclare{book}{fowler2010domain}
\bibitem{fowler2010domain}
\bibinfo{author}{Martin \surnamestart Fowler\surnameend}
  (\bibinfo{year}{2010}): \emph{\bibinfo{title}{Domain-specific languages}}.
\newblock \bibinfo{publisher}{Addison-Wesley Professional}.

\bibitemdeclare{misc}{ttcPN2SC}
\bibitem{ttcPN2SC}
\bibinfo{author}{Pieter~Van \surnamestart Gorp\surnameend} \&
  \bibinfo{author}{Louis~M. \surnamestart Rose\surnameend}
  (\bibinfo{year}{2013}): \emph{\bibinfo{title}{{The Petri-Nets to Statecharts
  Transformation Case}}}.
\newblock
  \bibinfo{howpublished}{\url{http://planet-sl.org/ttc2013/images/userdirs/122/ttc2013/pn2sc.pdf}}.

\bibitemdeclare{article}{sendall2003model}
\bibitem{sendall2003model}
\bibinfo{author}{Shane \surnamestart Sendall\surnameend} \&
  \bibinfo{author}{Wojtek \surnamestart Kozaczynski\surnameend}
  (\bibinfo{year}{2003}): \emph{\bibinfo{title}{Model transformation: The heart
  and soul of model-driven software development}}.
\newblock {\sl \bibinfo{journal}{Software, IEEE}}
  \bibinfo{volume}{20}(\bibinfo{number}{5}), pp. \bibinfo{pages}{42--45},
  \doi{10.1109/MS.2003.1231150}.

\end{thebibliography}

\appendix

\section{Appendix}

\begin{table}[ht]
	\centering
		\begin{tabular}{|l|c|c|c|}
		\hline
		Test case & Reading input & Transformation & Writing output \\
		\hline
		\hline
			sp200 & $12.96$ms & $8.78$ms & $8.62$ms \\
			\hline
sp300 & $19.87$ms & $13.06$ms & $12.72$ms \\
\hline
sp400 & $27.50$ms & $15.66$ms & $18.44$ms \\
\hline
sp500 & $32.80$ms & $21.14$ms & $21.62$ms \\
\hline
sp1000 & $72.09$ms & $55.31$ms & $41.10$ms \\
\hline
sp2000 & $176.91$ms & $111.13$ms & $103.97$ms \\
\hline
sp3000 & $295.12$ms & $175.32$ms & $142.97$ms \\
\hline
sp4000 & $465.48$ms & $248.64$ms & $160.39$ms \\
\hline
sp5000 & $618.60$ms & $299.40$ms & $241.14$ms \\
\hline
sp10000 & $1,887$ms & $745.50$ms & $499.84$ms \\
\hline
sp20000 & $6,273$ms & $1,549$ms & $1,600$ms \\
\hline
sp40000 & $22,012$ms & $2,923$ms & $2,277$ms \\
\hline
sp80000 & $83,417$ms & $5,639$ms & $6,453$ms \\
\hline
sp100000 & $128,445$ms & $7,040$ms & $7,343$ms \\
\hline
sp200000 & $488,709$ms & $13,532$ms & $18,317$ms \\
\hline
		\end{tabular}
	\caption{Execution times of the performance test cases}
	\label{tab:ExecutionTimesOfThePerformanceTestCases}
\end{table}

\end{document}